# Electro- and photoswitching of undulation structures in planar cholesteric layers aligned by a polyimide film possessing various values of the anchoring energy

Igor Gvozdovskyy

*Department of Optical Quantum Electronics, Institute of Physics of the National Academy of Sciences of Ukraine, Kyiv, Ukraine*

Institute of Physics of the National Academy of Sciences of Ukraine, Prospekt Nauki 46, Kyiv-28, 03680, Ukraine, telephone number: +380 44 5250862, *E-mail: igvozd@gmail.com

# Electro- and photoswitching of undulation structures in planar cholesteric layers aligned by a polyimide film possessing various values of the anchoring energy

In the present manuscript, a planar layer of the cholesteric liquid crystal, based on the photosensitive chiral dopant 2-(4'-phenylbenzylidene)-p-menthane-3-one dissolved in the nematic host E7, under the alternating electric field was explored. Due to the decrease in a ratio between the LC cell thickness $d$ and the cholesteric helical pitch $P$, different types of undulation instabilities, observed under the alternating electrical field, were switched by UV exposure. In particular, the influence of UV exposure and the value of the azimuthal anchoring energy on electro- and photoswitching of undulation instabilities (Helfrich-Hurault effect) was considered. The change in the ratio $d/P$ was realized owing to the unwinding of the photosensitive cholesteric helix during UV exposure in the LC cell with a fixed thickness $d$. Sequential transitions of types of the undulation instabilities from 2D to two 1D undulation structures ($1D_\parallel$ and $1D_\perp$) were achieved. The undulation structure $1D_\parallel$ with a controlled period of a grating in alternating electric field was examined. Dependencies of jumps of the period and the diffraction efficiency of a photosensitive cholesteric grating on the anchoring energy of alignment layers were analyzed.

The base for this manuscript was laid down by the studies carried out in the article by B. Senyuk *et. al.* "Elecrtically-controlled two-dimensional gratings based on layers undulations in cholesteric liquid crystals" Liquid Crystals IX, edited by Iam-Choon Khoo, Proceedings of SPIE Vol. 5936(SPIE, Bellingham, WA, 2005).

**Introduction**

Cholesteric liquid crystals (CLCs), due to their periodic helical structure, can selectively reflect light (the so-called phenomenon of the selective reflection or Bragg diffraction at the maximum wavelength $\lambda_{max} = \tilde{n} \cdot P$, where $\tilde{n}$ and $P$ are the average refractivity index and of the cholesteric helical pitch, respectively). Bragg diffraction is observed for the *planar* texture of cholesteric layers, where the helical axis is perpendicular to the plane of the LC cell. [1,2] When the helical axis is parallel to substrates of the LC cell, the

*homeotropic* texture of CLCs (or the so-called "fingerprint" nonuniform texture) can be obtained on a condition that the chiral torque is strong enough with respect to the elastic torque determined by the orientational elasticity and anchoring. [1-3] It is well known that a helix pitch *P* can be sensitive to various external fields (such as temperature, [4,5] electric [6-10] and magnetic [11,12] fields and light. [13-21] Due to the variation of the structural periodicity of CLCs, both planar and homeotropic textures give a chance for different optical and electro-optical applications, as for instance, light modulators, [22-24] reflectors, [25] optical data storage [26] and displays. [27,28]

In the last decade, a great deal of attention has been given to the study and application of the homeotropic texture of CLCs, controlled by both electric field and light. [9,14-17,19-21,29-39]

In the case of the uniform "fingerprint" texture [3,29-39], cholesteric diffraction gratings of the Raman-Nath type can be observed on a condition when the ratio $\lambda \cdot d/\Lambda^2 \ll 1$ (where, $\lambda$ is the light wavelength of the incident beam, $d$ and $\Lambda$ is the thickness and period of a grating, respectively). [40-42] It is known that the application of the alternating electric field, which is parallel to the helical axis of the planar texture of CLCs, gives rise to the formation of highly uniform structures (Helfrich-Hurault effect or the so-called undulation instabilities). [12,43,44] In addition, it should be emphasized that diffraction gratings of the Raman-Nath type were recently obtained for liquid crystal bubbles in the alternating electric field, by using the homeotropically alignment of CLCs, based on a nematic host with negative dielectric anisotropy ($\Delta\varepsilon < 0$). [24,45]

As was early mentioned [29], the varying uniform of the cholesteric periodicity can get complicated due to surface alignment problems. [30,31] For the strong surface anchoring energy *W*, when voltage *U* across the LC cell reaches some threshold value $U_{th}$, the cholesteric layers reorient parallel to the external field and undergo the sinusoidal periodic tilt. Such periodic deformation, when the cholesteric layers and the helical axis are tilted from their original orientation, was described in [43] by using angle *ψ* and the increment of tilt. Due to the competition between the free molecular rotation (the effect of "permeation" [44]) and the surface anchoring energy *W*, the maximal value of tilt is observed in the middle of the LC cell while minimal tilt is at the surface of substrates. [43,44] On the basis of the formation of undulation instabilities (or undulation structures), two types of periodic one-dimensional (1D) structures were classified: (a) the so-called developable-modulation (DM) and (b) growing-modulation (GM), depending on the ratio between the thickness *d* of the LC cell and cholesteric helical pitch $P$. [29,38,46] It was shown that for the ratio range $0.5 < d/P < 1$ the DM type of the cholesteric grating is formed with periodic one-dimensional structures, which are perpendicular to the rubbing direction (1D⊥), and their contrast increases with time. For the ratio $1.5 \leq d/P < 2.5$ two types of patterns (DM and GM) are formed simultaneously, but the GM type dominates as shown in. [29-31,38,46] The GM type of patterns is one-dimensional structures, which are parallel to the rubbing direction (1D∥). The usage of the modulated periodic undulation structure for switchable cholesteric gratings of the Raman-Nath type [40-42] with the electrically controlled period of stripe patterns, the dynamics of the pattern formation and the beam-steering characteristics of diffraction gratings were described in. [29,46,38] Recently, photo-switchable cholesteric gratings without applied electric field was studied in. [35]

It was also shown that when the electric field, thermal treatment or mechanical deformation was applied to a planar layer of CLC with $\Delta\varepsilon > 0$, then regions with two-dimensional periodic structures (the so-called square grid of lines, 2D undulation structure) appeared. [47] It was seen that 2D undulation structure consists of two 1D (1D∥ and 1D⊥) structures. According to theory [43], 2D undulation structure has perpendicular wave vectors, and they are independent of each other even when they have a small period. It was found that 2D undulation structure can appear in case of a large value of the ratio $d/P > 3$. [43,47] Recently, 2D undulation structure in the LC cell with finite surface anchoring energy by using fluorescence confocal polarizing microscope (FCPM) were assiduously studied. [48] The main diffraction features of the electro-switchable 2D grating based on the appearance of the Helfrich-Hurault effect [30,31] and bubbles [24] were considered.

In this manuscript, the idea was to create electro- and photocontrolled diffraction gratings due to sequential transitions between 2D and two 1D (1D∥ and 1D⊥) undulation structure, and it was achieved by using the photosensitive CLC layers of the fixed thickness $d$. The main characteristics of the two- and one-dimensional undulation structure for alignment films with various values of anchoring energy were studied. This work can serve as a model for the further usage of the photoreversible chiral dopants, as studied in, [13-17] having good spectral features (*i.e.* when absorption spectra of photoisomers are within various wavelength range). The findings described in this manuscript can be practically applied in optics to create 2D and 1D diffraction gratings with a fixed thickness $d$ of the cholesteric layer, *i.e.* all in one, that has been proposed here for the first time.

**Experiment**

To study an electrically switchable photosensitive cholesteric, the chiral nematic mixture with a long helical pitch was prepared.

As a photosensitive chiral dopant (ChD), the chemical compound 2-(4'-phenylbenzylidene)-p-menthane-3-one (PBM) synthesized and studied by L. Kutulya (Institute of Single Crystals of NAS of Ukraine), was used. [49,50]

Phototransformation of the PBM molecules is shown in Figure 1. UV spectra of *trans-* (or *E-*) and *cis-* (or *Z-*) isomers absorb in the wavelength range 227 ÷ 416 nm, as was early studied in. [50] Maximum absorbance of *E-* and *Z-*isomers is at the wavelength 298.5 and 285.7 nm, respectively. Due to a small shift of the absorption spectra of the isomers during UV exposure, a reversible photochemical reaction of the PBM molecules takes place but kinetically the rate of the $E \rightarrow Z$ process is about 1.5 ÷ 2.2 times greater than that for $Z \rightarrow E$. It can be roughly assumed that during UV exposure of the liquid crystalline mixture with PBM molecules by using a UV lamp with $\lambda_{max}$ = 365 nm the process of $E \rightarrow Z$ (or the so-called *trans-cis*-photoisomerisation) will occur. Quantum yields of photoisomerisation of PBM molecules, for instance in *n-*octane, [50] are shown in Figure 1. Owing to the small quantum yield of $Z \rightarrow E$ in the liquid crystal matrix [50] and strong coincidence of the absorption spectra of *Z* and *E* isomers, it can be concluded that predominantly the non-reversible $Z \rightarrow E$ process takes place, as was recently considered in. [20,35]

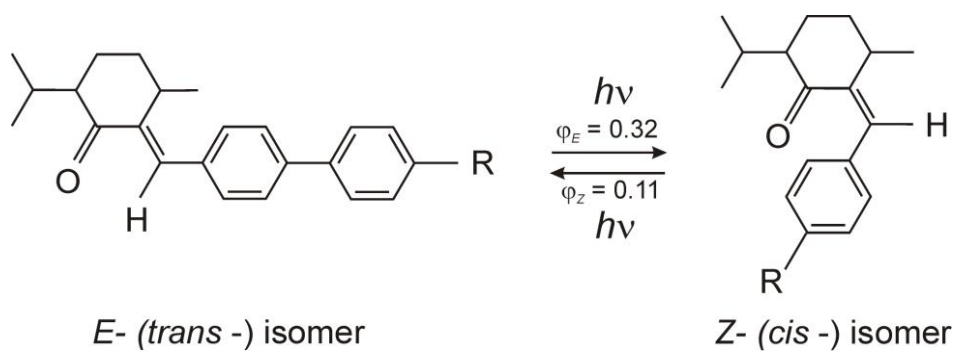

Figure 1. Reversible phototransformation of PBM molecules during UV irradiation in *n-*octane.

To prepare the left-handed CLC, the nematic liquid crystal E7, obtained by Licrystal, Merck (Darmstadt, Germany) was doped with a ChD PBM with concentration $C \sim 1$ wt.-%. The optical and dielectrical anisotropy of the nematic E7 at T = 20 °C, λ = 589.3 nm, and $f$ = 1kHz are $\Delta n$ = 0.225 ($n_e$ = 1.7472, $n_o$ = 1.5217) and $\Delta\varepsilon$ = +13.8, respectively. Splay, twist and bend elastic constants of nematic E7 are $K_{11}$ = 11.7 pN, $K_{22}$ = 6.8 pN, $K_{33}$ = 17.8 pN, respectively. [51-53]

To measure cholesteric pitch *P,* the well-known Grandjean-Cano method with the preparing of the wedge-like LC cell was used. [54,55] The cholesteric helical pitch was about 4 µm.

To obtain the planar alignment of the nematic liquid crystal E7 and CLC on his base, both the polyimide PI2555 (HD MicroSystems, USA) and 2-% DMF solution of oxidianiline-polyimide (ODAPI) (Kapton synthesized by I. Gerus, Institute of Bio-organic Chemistry and Petrochemistry, NAS of Ukraine) were used. The polyimide PI2555 and ODAPI films were deposited on glass substrates by the spin-coating method (6000 rpm, 30 s). The polyimide PI2555 films (reference substrates) were annealed at 180$^o$ C for 30 min, while ODAPI films (tested substrates) were annealed at 190$^o$ C for 90 min. To obtain the CLC planar alignment, the ODAPI films were also spin-coated on indium-tin oxide (ITO)-coated glass substrates, which were prepared in V. Ye. Lashkaryov Institute of Semiconductor Physics NAS of Ukraine, with further etched ITO. Thereafter thin polyimide films were unidirectionally rubbed.

In order to get alignment ODAPI layers having various values of anchoring energy *W,* a different number of times of unidirectional rubbings $N_{rubb}$ were made. To measure the anchoring energy the idea of the combined twist LC cell was used. [56] The twist LC cell consists, on the one hand, of the tested plate, coated by the ODAPI film with different $N_{rubb}$ =1 ÷ 25 and the pressure of rubbing about 2100 N/m$^2$, and, on the other hand, of the reference plate, coated by polyimide PI2555 having strong anchoring energy $W$ = (4 ± 1)×10$^{-4}$ J/m$^2$ for $N_{rubb}$ = 7 ÷ 15 with the pressure of rubbing in range 800 ÷ 850 N/m$^2$, as was recently shown in. [30,48,57] The easy axis of the tested (the rubbing ODAPI film with different $N_{rubb}$) and reference (rubbing polyimide PI2555 with $N_{rubb}$ = 10) substrates is given by the directions of rubbing on the one hand, a 45$^o$ angle to the selected side of the tested substrate and on the other hand, along the selected side of the reference substrate. In this case the angle $\varphi_0$ between the easy axis of the reference and the tested substrate was 45 degree. The thickness *d* of the plane-parallel twist LC cells was around 20.2 µm. The LC cells were filled with the nematic liquid crystal E7 in the isotropic phase ($T_{iso}$ = 59.8 $^o$C) [51] and slowly cooled to the room temperature to avoid the possible flow alignment.

To study the influence of UV exposure on the unwinding of the cholesteric pitch $P$, the wedge-like LC cells, assembled with a pair of substrates having the same $N_{rubb}$, were used. The thickness $\Delta d$ of the wedge-like LC cells were in range $20.5 \pm 0.3$ µm.

The plane-parallel symmetrical LC cells with the unidirectionally rubbed ODAPI films were assembled using two substrates, which have the opposite rubbing on both aligning surfaces. All LC cells were assembled with a thickness set by Mylar spacers and controlled by the interference method measuring the transmission spectrum of the empty cell. Thicknesses of LC cells were in range $d = 20.5 \pm 0.3$ µm.

To avoid various defects for a planar cholesteric structure the LC cell was filled with the chiral nematic mixture in the isotropic phase and slowly (~ $0.1°$ C/min) cooled as described elsewhere. [30]

In order to avoid the edge effect during the formation of undulation structure of CLC under the AC field (the appearance of oil streaks through the LC cell) the glass substrates with ITO films were etched with the square area approximately $5 \times 5$ mm$^2$, as it was recently shown in. [30,56]

To form undulation structure in the planar CLC an AC voltage (frequency f = 10 kHz) was applied across the sample by using a frequency generator GZ-109 (Pskov Region, Russia).

The 2D and 1D undulation structures formed in the AC field were viewed through a polarising microscope BioLar (PZO, Warszawa, Poland) equipped with a digital camera Nikon D80 (Japan).

The illumination of LC cell with the chiral mixture was carried out by a UV lamp ($\lambda_{max} = 365$ nm) with its total power about 6 W.

To measure the twist angle $\varphi$ and the intensity $I_m$ of diffraction orders ($m = 0, \pm 1, \pm 2,...$) the output monochromatic linear-polarized light (TEM$_{00}$) with $\lambda = 632.8$ nm and its power about 1.5 mW from He-Ne laser LGN-207a (Lviv, Ukraine) and a silicon photodiode (PD) FD-18K (Kyiv, Ukraine) with the spectral range $470 \div 1100$ nm and maximal sensitivity in range $850 \div 920$ nm were used. PD was connected to the oscilloscope Hewlett Packard 54602B 150MHz (USA).

**Results and discussion**

*3.1. Measurement of the azimuthal anchoring energy W of ODAPI films with different $N_{rubb}$*

First of all, the dependence of the azimuthal anchoring energy $W$ of ODAPI films on a different number of times of the rubbing $N_{rubb}$ was examined.

An experimental setup for the measurement of the twist angle $\varphi$ of the combined twist LC cells is shown in Figure 2 (a). The LC cells were placed between a pair of parallel polarizes (*i.e.* the polarization plane of polarizer (P) and analyzer (A) coincides), as shown in Figure 2 (a). The rubbing direction of the reference PI2555 substrate coincided with the polarization plane of a pair of polarizers, while the rubbing direction of the tested ODAPI substrate was at an angle $\varphi_0 = 45°$ to the polarization plane of P and A. It is seen that the linear-polarized incident beam passes from He-Ne laser through P and the LC cell toward A and further to the photodetector PD. In the twist LC cell the beam $\vec{E}_P$ is rotated at a certain twist angle $\varphi$, which depends on anchoring energy $W$ of the test substrate. [56] Behind the LC cell the rotated beam $\vec{E}_P^{LC}$ propagates toward the analyzer A, as can be seen in Figure 2 (a). Due to the different orientation of the polarization plane between the rotated beam $\vec{E}_P^{LC}$ (after it passed through the twist LC cell) and the polarization plane of the analyzer (A), a certain part $\vec{E}_A^{LC}$ of the beam $\vec{E}_P^{LC}$ will reach the photodetector PD. In order to measure a value of the twist angle $\varphi$, depending on anchoring energy $W$ [56] (or $N_{rubb}$) of the tested ODAPI film, the analyzer A was clockwise rotated through some angle $\varphi'$ to obtain the minimal intensity (*i.e.* $I \rightarrow I_{min}$) of the incident beam on PD connected to the oscilloscope. In this case the value of the angle $\varphi'_{I \rightarrow Imin}$ is the real twist angle $\varphi$ between the easy axis of the reference and tested substrates with rubbed PI2555 and ODAPI layers for certain experimental conditions.

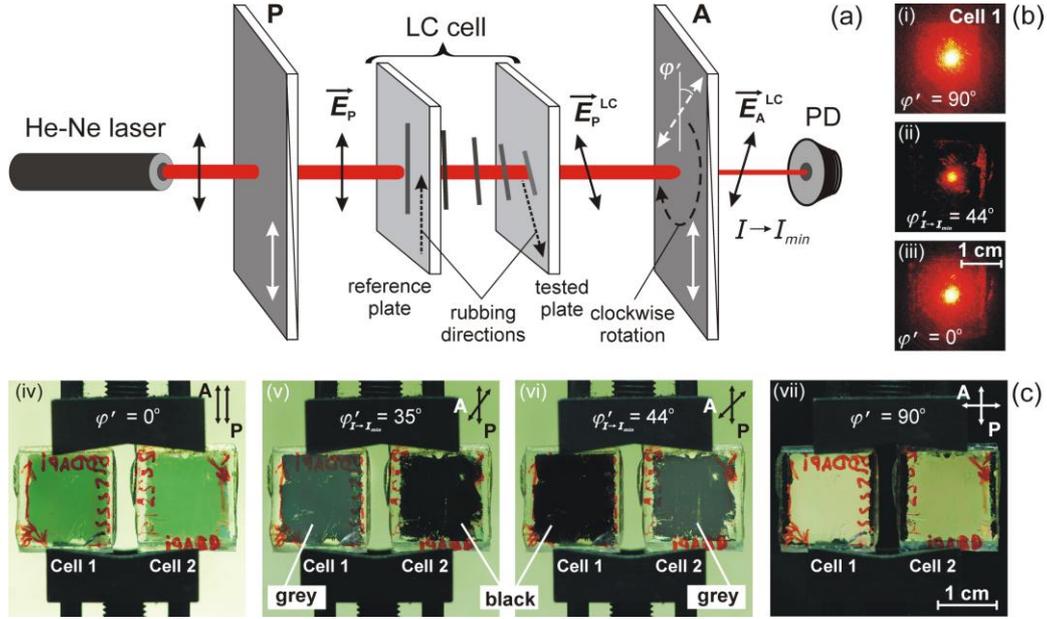

Figure 2. (a) Schema of the measurement of the twist angle $\varphi$ of the combined twist LC cell, assembled with the reference PI2555 ($N_{rubb} = 10$) and tested ODAPI substrates. (b) Photographs of the laser beam passing through the twist LC cell consisting of the reference PI2555 and tested ODAPI ($N_{rubb} = 10$) substrates placed between a pair of polarizers with a different angle of the plane polarization: (i) $\varphi' = 90°$, (ii) $\varphi' = 44°$ and (iii) $\varphi' = 0°$. (c) Photographs of two twist LC cells, placed between a pair of polarizers with a different angle of the plane polarization: (iv) $\varphi' = 0°$; (v) $\varphi' = 35°$; (vi) $\varphi' = 44°$ and (viii) $\varphi' = 90°$. Cell 1 consists of the tested substrate having anchoring energy $W = 1.3 \times 10^{-4}$ J/m$^2$. Cell 2 consists of the tested substrate having anchoring energy $W = 11 \times 10^{-6}$ J/m$^2$. The thickness of the twisted LC cells was $d = 20.2$ μm.

In Figure 2 (b) photographs of the intensity of the beam $\vec{E}_A^{LC}$, passing through the LC cell (for instance, strong anchoring energy $W = 1.3 \times 10^{-4}$ J/m$^2$) placed between a pair of polarizers are shown. For different angles of the polarization plane between P and A, namely $\varphi' = 90$, 44 and 0 degree, the LC cell is shown at photographs (i), (ii) and (iii), respectively. It is easily seen that when the analyzer is rotated on an angle of 44 degree the intensity of the laser beam is minimum $I_{min}$ (the small spot of the laser beam after passing through the LC cell, placed between a pair of polarizers, is observed).

In Figure 2 (c) photographs (iv), (v), (vi) and (vii) show two twist combined LC cells, assembled with the tested substrate, having both strong anchoring energy $W = 1.3 \times 10^{-4}$ J/m$^2$ (cell 1) and weak anchoring energy $W = 11 \times 10^{-6}$ J/m$^2$ (cell 2). At photographs the analyzer A was rotated through the fixed angle $\varphi' = 0$, 35, 44 and 90 degree. It is seen that light is not transmitted through the combined twist LC cell in two cases. Firstly, the LC cell 2 (see photograph (v)), having the tested substrate possessing weak anchoring energy $W = 11\times10^{-6}$ J/m$^2$ ($N_{rubb} = 1$), and the angle between the polarizers is about 35 degree, does not transmit light (the black zone), while the LC cell 1 partially transmits light (the grey zone). Secondly, the LC cell 1 (see photograph (vi)), consisting of the tested substrate having strong anchoring energy $W = 1.3\times10^{-4}$ J/m$^2$ ($N_{rubb} = 10$), and the angle between the polarizers is about 44 degree (no light is transmitted through the LC cell 1 (the black zone)), and *vice versa* the LC cell 2 partially transmits light (the grey zone).

For three different samples, the dependence of the twist angle $\varphi$ on the number of rubbings $N_{rubb}$ is shown in Figure 3(a). The average value of the twist angle $<\varphi>$ as a function of the $N_{rubb}$ is shown by diamond symbols (to ease, symbols were connected by the solid line). The maximal twist angle $<\varphi> \approx 44°$ was observed for $N_{rubb} = 10$ and 15, as can be seen from Figure 3(a).

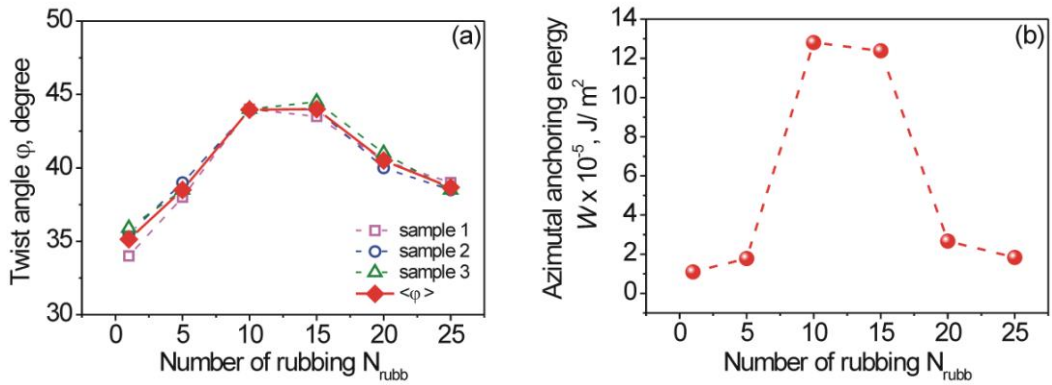

Figure 3. Dependencies of the twist angle $\varphi$ of different samples and their average twist angle $<\varphi>$ (a) and calculated azimuthal anchoring energy $W$ for the average twist angle $<\varphi>$ (b) on $N_{rubb}$ for the twist LC cells ($d = 20.2$ µm). LC cells were constructed from the reference substrate with rubbed PI2555 ($N_{rubb} = 10$) and the tested substrate with ODAPI film, which is rubbed with different $N_{rubb} = 1 \div 25$.

According to [56,59,60] the twist angle $\varphi$ is related to the azimuthal anchoring energy $W$ as:

$$W = K_{22} \cdot \frac{2\cdot\sin(\varphi)}{d\cdot\sin 2(\varphi_0 - \varphi)} \quad (1),$$

where $d$ is the thickness of the LC cell, $\varphi_0 = 45^o$ is the angle between the easy axis of the reference and tested substrate with rubbed PI2555 and ODAPI layer, respectively and $\varphi$ is the measured twist angle.

The dependence of the calculated azimuthal anchoring energy $W$ on different $N_{rubb}$ of the tested ODAPI layer, by using Equation (1), is shown in Figure 3(b) for the average value of the twist angle $<\varphi>$. For these experimental conditions, values of anchoring energy $W$ at different $N_{rubb}$ and pressure 2100 N/m$^2$ are also listed in Table 1.

Table 1. Calculated azimuthal anchoring energy $W$ and average twist angle $<\varphi>$ for different number of times of the rubbing $N_{rubb}$

| Number of times of the rubbing $N_{rubb}$ | Average twist angle $<\varphi>$ (degree) | Azimuthal anchoring energy $W$ (J/m$^2$) |
|---|---|---|
| 1 | 35.1 | $11 \times 10^{-6}$ |
| 5 | 38.5 | $17 \times 10^{-6}$ |
| 10 | 44 | $1.3 \times 10^{-4}$ |
| 15 | 43.9 | $1.2 \times 10^{-4}$ |
| 20 | 40.5 | $27 \times 10^{-6}$ |
| 25 | 38.6 | $18 \times 10^{-6}$ |

*3.2. Unwinding of the helix of the photosensitive CLC under UV exposure of different LC cells with alignment ODAPI films having certain value of anchoring energy*

At the beginning, the changes in the length of the cholesteric pitch $P$ during UV exposure of the wedge-like LC cells were studied. After each procedure of UV exposure (at the certain exposure time $t_{irr}$) the cholesteric pitch $P$ was calculated by using the well-known equation: [54,55]

$$P = 2 \cdot \Delta d / N_{GC} \quad (2),$$

where $\Delta d$ and $N_{GC}$ is the thickness of the wedge-like LC cell and number of Grandjaen-Cano stripes, respectively.

The dependence of the cholesteric pitch $P$, calculated by Equation (2), on UV exposure for ODAPI films possessing various values of anchoring energy $W$ (or $N_{rubb}$) are shown in Figure 4. Owing to *trans-cis* photoisomerisation of PBM molecules during UV exposure of CLC, the increase in the concentration of the *cis*-isomer leads to the decrease in the helical twisting power ($\beta$ or HTP) of the cholesteric mixture. [50] It is well known that the chiral power $\beta$ and the cholesteric pitch $P$ are related by the equation $\beta = (C \times P)^{-1}$. [1,2]

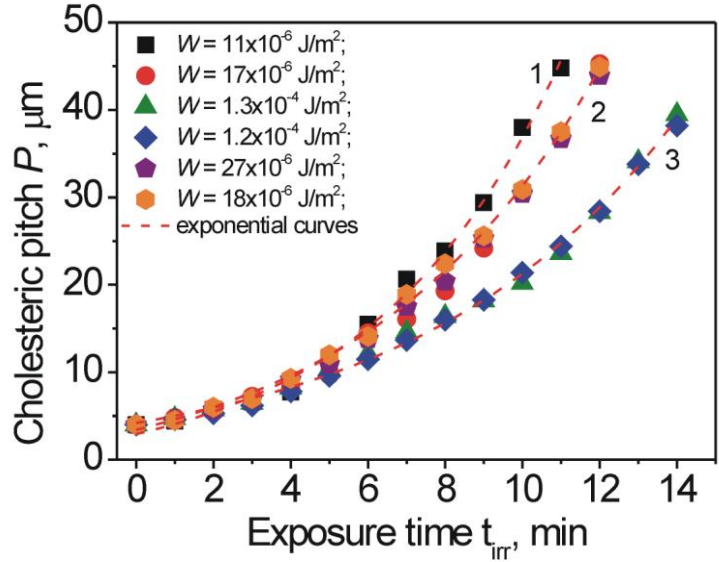

Figure 4. Dependence of the cholesteric pitch $P$ on the exposure time $t_{irr}$ in the wedge-like LC cells consisting of ODAPI films possessing various values of anchoring energy $W$: (solid squares, curve 1) – $11\times10^{-6}$ J/m$^2$ ($N_{rubb}$ = 1), $\Delta d$ = 20.2 μm; (solid circles, curve 2) – $17\times10^{-6}$ J/m$^2$ ($N_{rubb}$ = 5), $\Delta d$ = 20.8 μm; (solid triangles, curve 3) – $1.3\times10^{-4}$ J/m$^2$ ($N_{rubb}$ = 10), $\Delta d$ = 20.8 μm; (solid diamonds, curve 3) – $1.2\times10^{-4}$ J/m$^2$ ($N_{rubb}$ = 15), $\Delta d$ = 20.5 μm; (solid pentagons, curve 2) – $27\times10^{-6}$ J/m$^2$ ($N_{rubb}$ = 20), $\Delta d$ = 20.4 μm; and (solid hexagons, curve 2) – $18\times10^{-6}$ J/m$^2$ ($N_{rubb}$ = 25), $\Delta d$ = 20.2 μm.

As can be seen from Figure 4, during UV exposure of the chiral mixture the cholesteric helical pitch $P$ is increased. Due attention should be given to the fact that for different wedge-like LC cells, which are distinguished by substrates possessing certain value of anchoring energy $W$ (or $N_{rubb}$), temporal dependences of the pitch $P$ during UV exposure differ from one another. It is easily seen that the increase in anchoring energy $W$ leads to the decreasing changes in the pitch $P$ with the exposure time $t_{irr}$ (or, in other words, the unwinding speed of the cholesteric helix is decreased). When anchoring energy $W$ is moderately different (for instance, curve 2: $W$ = 17, 27 and $18\times10^{-6}$ J/m$^2$ at $N_{rubb}$ = 5, 20 and 25, respectively or curve 3: $W$ = 1.3 and $1.2\times10^{-4}$ J/m$^2$ at $N_{rubb}$ = 10 and 15, respectively), the unwinding speeds of the cholesteric helixes are about the same. In the case of the weak anchoring energy $W$ = $11\times10^{-6}$ J/m$^2$ (at $N_{rubb}$ = 1), the change in the helix pitch $P$ with the exposure time $t_{irr}$ is maximal (Figure 4, curve 1). It is obvious that under certain experimental conditions (namely, during the same exposure time $t_{irr}$), substrates having strong anchoring energy $W$ can heavily hold in place without the unwinding of the cholesteric helix, though UV exposed, than for substrates possessing weak anchoring energy $W$. As can be seen form Figure 4, with prolonged UV irradiation, when the concentration of the *cis*-isomer of the PBM molecules is increased (*i.e.* $\beta$ is reduced), the difference between the unwinding speeds of the cholesteric helix for the weak and strong anchoring energy is strongly distinct.

### *3.3. Photoswitching of undulation structure for different LC cells with alignment ODAPI films having various values of anchoring energy*

It is well known that the appearance of undulation structure depends on the ratio between the thickness $d$ of the LC cell and the cholesteric pitch $P$. [3,14,30,43,44,48]

To study the photoswitching of undulation structure from 2D to 1D the CLC with the initial cholesteric pitch $P \approx 4$ μm before UV irradiation was taken. The thickness of the LC cells was in range of $d = 20.5 \pm 0.3$ μm. At the beginning, the ratios between the thickness of the LC cell and the cholesteric pitch were in range $d/P \sim 5 \div 5.2$. For these ratios $d/P$ the LC cells with uniform planar cholesteric textures without any defects were observed. With an AC electric field the liquid crystal molecules are oriented along the applied field. At a certain threshold of voltage $U_{th}$, due to the competition between the dielectric and surface energies, 2D undulations are detected.

As was mentioned above, owing to the phototransformation of the PBM molecules (Figure 1), the cholesteric pitch $P$ is increased during UV exposure of the LC cell. The change in the cholesteric pitch $P$ for the fixed thickness of the LC cell $d$ leads to the decrease in the ratio $d/P$. The dependences of the ratio $d/P$ on the exposure time $t_{irr}$ for different LC cells, consisting of substrates having certain value anchoring energy $W$, are shown in Figure 5. By taking into account small variations of the thickness $d$, it can be concluded that the dependence $d/P(t_{irr})$ is not identical for the LC cells, which are different by anchoring energy $W$ (or $N_{rubb}$) of ODAPI films. As can be seen from the inset in Figure 5, dependencies of the ratio $d/P(t_{irr})$ presented on a logarithmic scale are essentially distinct. It is seen that for the weak anchoring energy $W$ (curve 1 or curve 2) the stable existence of undulation structure is observed during shorter UV exposure and vice versa in the case of the strong anchoring (curve 3).

Figure 5 shows ranges of the appearance and the stable existence of different undulation structures at the certain ratio $d/P$, having come about through UV exposure of the sample. For instance, under certain experimental conditions stable 2D undulation structure is observed within the range of ratios ~5.2 ÷ 3.6, while stable parallel $1D_\parallel$ and orthogonal $1D_\perp$ undulation structure to the rubbing direction are within the ratio range ~3.5 ÷ 1.2 and ~1.1 ÷ 0.4, respectively.

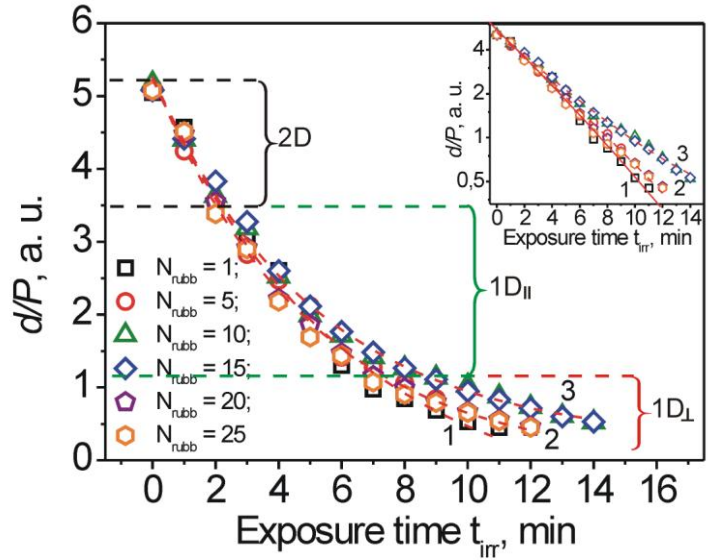

Figure 5. Dependence of the ratio $d/P$ on the exposure time $t_{irr}$ in the plane-parallel symmetrical LC cells consisting of ODAPI films possessing various values of anchoring energy $W$: (open squares, curve 1) – $11\times10^{-6}$ J/m$^2$ ($N_{rubb}$ = 1), $d$ = 20.2 μm; (open circles, curve 2) – $17\times10^{-6}$ J/m$^2$ ($N_{rubb}$ = 5), $d$ = 20.5 μm; (open triangles, curve 3) – $1.3\times10^{-4}$ J/m$^2$ ($N_{rubb}$ = 10), $d$ = 20.7 μm; (open diamonds, curve 3) – $1.2\times10^{-4}$ J/m$^2$ ($N_{rubb}$ = 15), $d$ = 20.3 μm; (open pentagons, curve 2) – $27\times10^{-6}$ J/m$^2$ ($N_{rubb}$ = 20), $d$ = 20.3 μm; and (open hexagons, curve 2) – $18\times10^{-6}$ J/m$^2$ ($N_{rubb}$ = 25), $d$ = 20.3 μm. The inset depicts the dependence of $d/P(t_{irr})$ in a logarithmic scale.

Owing to the unwinding of the helix of the photosensitive cholesteric layer with a constant thickness $d$, the changes in the ratio of $d/P$ during UV exposure were realized. The fact that the ratio $d/P$ is controlled by UV radiation can be useful for the formation of 2D and 1D undulation structures and their transitions (or, in other words, the so-called photoswitching) in the electric field. Furthermore, depending on the exposure time $t_{irr}$ for 1D undulations, formed by the AC electric field, either orthogonal (⊥) or parallel (∥) orientation with respect to the rubbing direction of alignment films can be observed. As a result, photoswitching between $1D_\parallel$ and $1D_\perp$ (or a 90 degree rotation of $1D_\parallel$ undulation structure) is achieved. In Figures 6 (a), (b) and (c), sequential appearances of different types of undulation structure, formed in the AC electric field after a certain exposure time $t_{irr}$ of the LC cell with a fixed thickness $d = 20.7$ µm, are shown. As an example, the diffraction patterns for 2D and two 1D ($1D_\parallel$ and $1D_\perp$) undulation structures are shown in Figures 6 (d), (e) and (f), respectively.

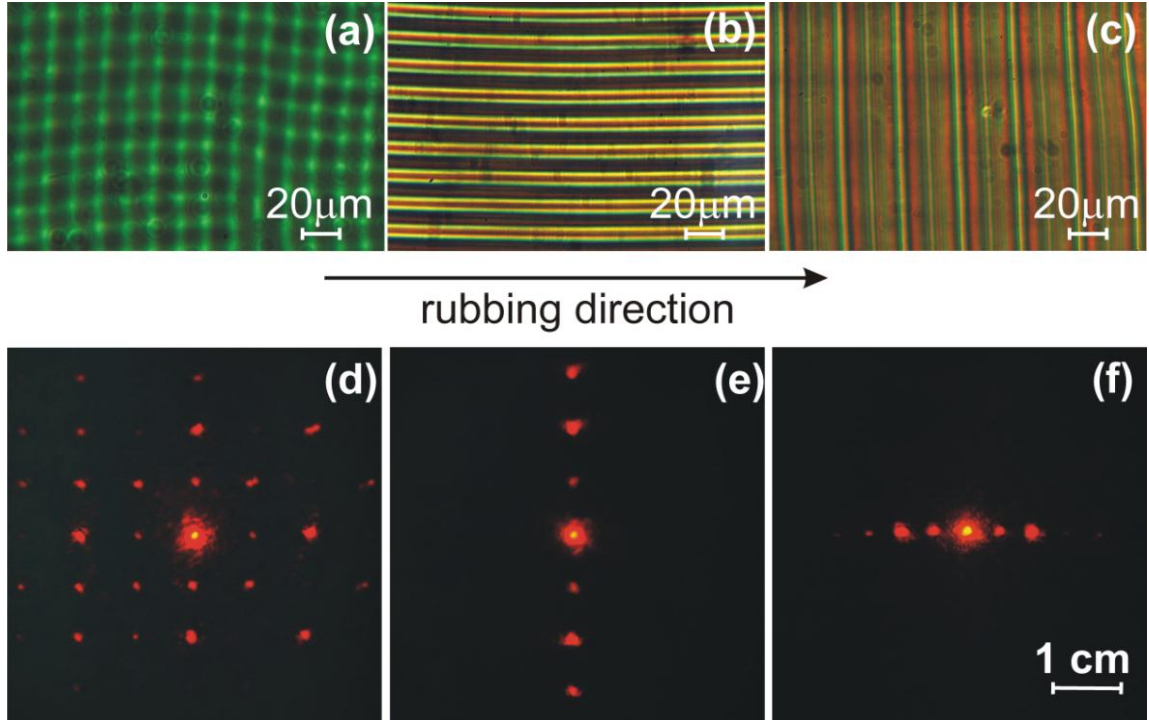

Figure 6. Photographs of undulation structure formed in the AC electric field ($f = 10$ kHz) at the threshold voltage $U_{th}$ and exposure time $t_{irr}$: (a) 2D structure with period of grating $\Lambda = 13$ µm at $t_{irr} = 1$ min, $d/P \approx 4.4$ and $U_{th} = 5.7$ V; (b) $1D_\parallel$ structure with period of grating $\Lambda = 13.5$ µm at $t_{irr} = 4$ min, $d/P \approx 2.5$ and $U_{th} = 4.2$ V; and (c) $1D_\perp$ structure with period of grating $\Lambda = 29.5$ µm at $t_{irr} = 11$ min, $d/P \approx 0.9$ and $U_{th} = 2.4$ V. Photographs of the diffraction pattern for: (d) 2D, (e) $1D_\parallel$ and (f) $1D_\perp$ undulation structure. The distance between the LC cell and viewing screen of the diffraction pattern was about 16 cm. The LC cell consists of a pair of substrates with ODAPI films having strong anchoring energy $W = 1.3 \times 10^{-4}$ J/m$^2$ ($N_{rubb} = 10$). The thickness of the plane-parallel symmetrical LC cell was 20.7 µm.

For alignment ODAPI films possessing various values of anchoring energy $W$, the dependence of the period $\Lambda$ of undulation structure (in other words, the period of grating) on the exposure time $t_{irr}$ is presented in Figure 7. After each UV exposure, the unwinding of the cholesteric helix leads to the increase in the period $\Lambda$ of undulation structure, formed in the LC cell under the alternating electric field application. In Figure 7 each point corresponds to period of grating (or $\Lambda$) under certain experimental conditions, namely: a certain exposure time $t_{irr}$ and threshold voltage $U_{th}$, which corresponds to the value of voltage when undulation structure are appearing.

It is seen that both for weak $W = 11\times10^{-6}$ J/m$^2$ (Figure 7 (a)) and strong anchoring energy $W = 1.3\times10^{-4}$ J/m$^2$ (Figure 7 (b)), there are always three different branches of the dependence period of grating $\Lambda$ on the exposure time $t_{irr}$, that correspond to different undulation structure. Each branch concerns a certain range of ratios $d/P$ as shown in Figure 5. During UV exposure of the LC cell the increase in period of grating is observed, as can be seen from Figure 7. In the case of 2D structures, period of grating increases under not a long exposure time $t_{irr}$. For both perpendicular and parallel directions regarding the rubbing direction of ODAPI films, period of grating changes uniformly in magnitude, though the difference between period of grating in two orthogonal directions can be less than 1 %, as was recently measured by using FCPM. [31]

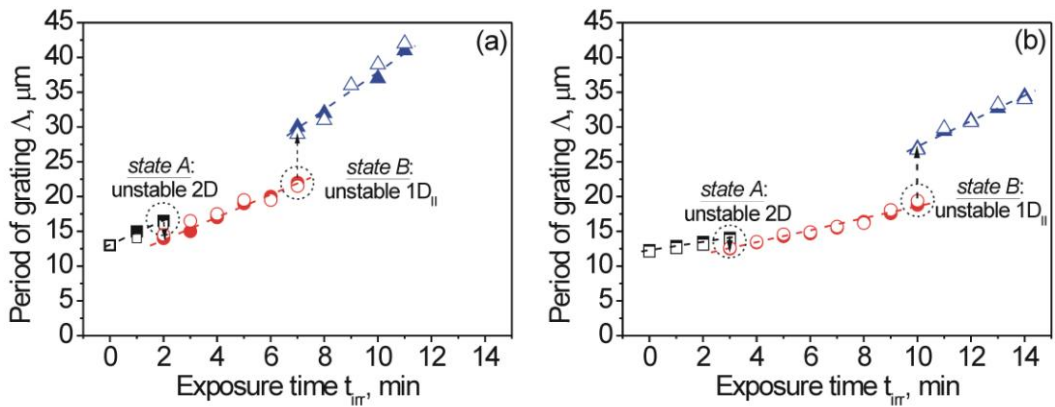

Figure 7. Dependence of period of grating $\Lambda$ on the exposure time $t_{irr}$ for the plan-parallel LC cells, assembled with a pair of substances with ODAPI films possessing: (a) weak $W = 11\times10^{-6}$ J/m$^2$ ($N_{rubb} = 1$) and (b) strong anchoring energy $W = 1.3\times10^{-4}$ J/m$^2$ ($N_{rubb} = 10$). Thicknesses of the LC cells were: (a) - 20.2 and 20.3 μm and (b) - 20.7 and 20.5 μm (solid and open symbols, respectively). Dashed circles show the undulation structure transitions: *state A* - 2D → 1D$_\parallel$ and *state B* - 1D$_\parallel$ → 1D$_\perp$, where 2D undulation structure – square symbols, 1D$_\parallel$ undulation structure – circle symbols and 1D$_\perp$ undulation structure – triangle symbols.

It should be noted that there are two states when undulation structure transitions occur. For *state A* the transition between 2D and 1D$_\parallel$ structures is observed. For *state B* the transition between 1D$_\parallel$ and 1D$_\perp$ structures takes place. In Figure 7 dashed circles show these states. In transition points, which correspond to a certain exposure time $t_{irr}$ and threshold voltage $U_{th}$, 2D and 1D$_\parallel$ structures are unstable with time. As can be seen from Figure 7, for both states the jumps of period of grating under certain conditions ($t_{irr}$ and $U_{th}$) are observed. However, for 2D → 1D$_\parallel$ structure transition, the jump of period of grating occurs by the decrease in the period of grating magnitude. It should be noted that after transition 2D → 1D$_\parallel$ the magnitude of period of grating is roughly equal to the magnitude of period of grating (2D structure) before UV exposure, as can be seen from Figure 7. However, in the case of 1D$_\parallel$ → 1D$_\perp$ transition the jump with the increase in period of grating is observed.

For substrates with ODAPI films, possessing strong anchoring energy $W = 1.3 \times 10^{-4}$ J/m$^2$ ($N_{rubb} = 10$), the change in period of grating with the exposure time $t_{irr}$ is less than for substrates having weak anchoring energy $W = 11 \times 10^{-6}$ J/m$^2$ ($N_{rubb} = 1$), as it can be easily seen from the tilt angle of each branch, by comparing Figure 7 (a) and Figure 7 (b). Inasmuch as HTP of ChD PBM with the UV exposure time changes alike in CLC, but the helix pitch unwinds in a different way for various values of $W$ (Figure 4), then it is obviously that the reason for the different tilt angles of dependences $\Lambda(t_{irr})$ may be in various values of anchoring energy $W$ of rubbed ODAPI layers. As can be seen from Figure 7, after each UV exposure, in the alternating electric field, the formation of undulation structure with large period of grating is observed for ODAPI films having weak anchoring energy (Figure 7 (a)).

The change of period of grating $\Lambda$ with the exposure time $t_{irr}$ is a non-linear dependence for each type of undulation structure. It is in good agreement with results obtained during studies of the cholesteric mixture [14] and polymer cholesteric grating during UV exposure. [33] As an example, let us consider in detail the dependence $\Lambda(t_{irr})$ for each type of undulation structure, which are shown in Figure 8. Two LC cells were assembled using a pair of substrates having strong anchoring energy ($W = 1.2 \times 10^{-4}$ J/m$^2$, $N_{rubb} = 15$). For both cells the thickness of CLC layers was 20.3 μm (open and solid symbols). As can be seen from Figure 8, owing to *trans-cis* photoisomerisation of the dopant PBM (Figure 1), the unwinding of the helical structure leads to the increase in period of grating, which can be approximated by a certain exponential curve $y(x) \sim y_0 + A(1 - e^{-x/B})$, where A and B are some fitting constants.

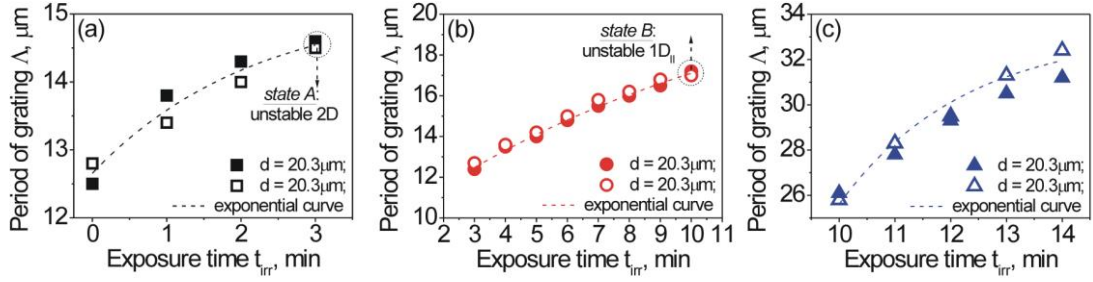

Figure 8. Dependence of period of grating Λ on the exposure time $t_{irr}$ for undulation structure: (a) 2D, (b) 1D$_\parallel$ and (c) 1D$_\perp$. The LC cells consist of two substrates having strong anchoring energy $W = 1.2 \times 10^{-4}$ J/m$^2$ ($N_{rubb} = 15$). Thicknesses of LC cells (open and solid symbols) were 20.3 μm. Dashed lines are exponential curves with some constants A and B.

As was mentioned above, under certain experimental conditions (for a fixed exposure time $t_{irr}$ and a certain value of the threshold voltage $U_{th}$) the appearance of undulation structure is observed. Mainly, it depends on the threshold voltage $U_{th}$. The undulation structure type strongly depends on the ratio value of $d/P$, usually changed by using different thickness of the LC cell at the fixed cholesteric helical pitch $P$. [30,31,37,47,48] As it was realized in, [24,33,45] here, the ratio $d/P$ was also changed, owing to the unwinding of the cholesteric helix, during UV exposure of the LC cell with a fixed thickness $d$. It was experimentally found that after each sequential UV exposure the appearance of undulation structure is observed at a smaller threshold voltage $U_{th}$. The unwinding of the cholesteric helix leads to the decrease in the threshold voltage $U_{th}$. The dependence of the threshold voltage $U_{th}$ on the various values of ratio $d/P$ is a linear function, as can be seen from Figure 9.

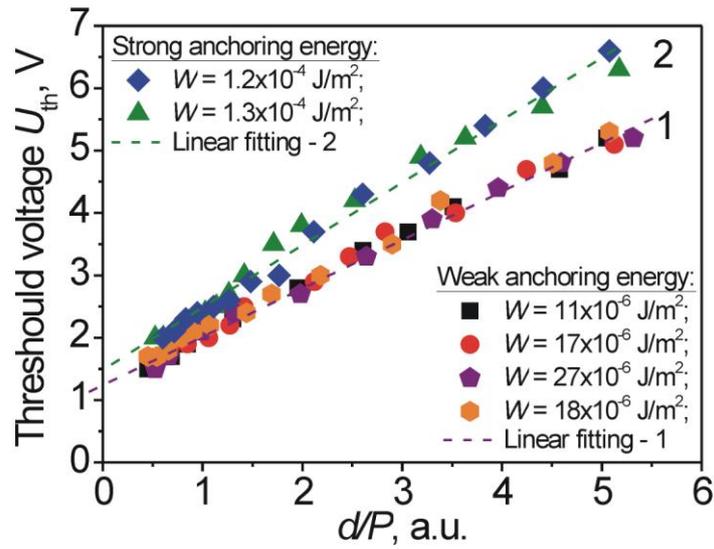

Figure 9. Dependence of the threshold voltage $U_{th}$ on the ratio $d/P$ for a range of weak anchoring energy $W \sim (11 \div 27) \times 10^{-6}$ J/m$^2$ (curve 1: solid squares, circles, pentagons, and hexagons) and strong anchoring energy $W \sim (1.2 \div 1.3) \times 10^{-4}$ J/m$^2$ (curve 2: solid diamonds and triangles).

In the case of weak anchoring energy $W$ (dashed line 1) of the alignment film there is a need to apply lower magnitude of the threshold voltage $U_{th}$ to observe the appearance of undulation structure (or, in other words, the deformation of cholesteric layers) and vice versa: for alignment film possessing strong anchoring energy $W$ (dashed line 2), there is need to use higher magnitude of $U_{th}$ to deform the cholesteric.

### *3.4. Jumps of period of grating of 1D$_\parallel$ structure in AC electric field for different LC cells with alignment ODAPI films having a certain value of anchoring energy*

This section is dedicated to the jumps of period of grating in the case of the appearance of 1D$_\parallel$ structures with alternating electric field for alignment ODAPI films having various values of anchoring energy $W$ as it was previously studied in. [61] However, it should be noted that the dependence of period of grating $\Lambda$ on the applied voltage for 2D undulation structure for various cholesteric mixtures were in detail observed in. [30,31]

As was mentioned above, before application of the electric field to a thin cholesteric layer, UV exposure of a photosensitive helix leads to the decrease in the ratio $d/P$. It is a reason for changes in the undulation structure type in consecutive order 2D → 1D$_\parallel$ and 1D$_\parallel$ → 1D$_\perp$ with the applied voltage $U$. It is well known [30,31,38,46] that period of grating can be switched by the electric field when 1D∥ structure is observed and ratio is $d/P > 1.5$, while for 1D⊥ structure ($d/P < 1$), changes in contrast of a diffraction pattern and no switching of period of grating with the applied voltage $U$ can be observed.

After a certain fixed exposure time $t_{irr}$ (for instance, 2, 4 and 6 min), in the case of ratio $d/P > 1.2$, when 1D∥ structure is appeared with voltage $U_{th}$ (for example, as shown in Figure 8 (b)), period of grating is switched by the electric field as shown for 20.3 μm LC cell, assembled with a pair of substrates having weak anchoring energy $W = 11 \times 10^{-6}$ J/m$^2$ (at $N_{rubb} = 1$) (Figure 10 (a) – (c)). For each exposure time $t_{irr} = 2$, 4 and 6 min the ratio of $d/P$ is about 3.5 (Figure 10 (a)), 2.6 (Figure 10 (b)) and 1.3 (Figure 10 (c)), respectively and jumps of period of grating of undulation structure in the alternating electric field with frequency $f = 10$ kHz are observed.

To compare the influence of anchoring energy $W$ on jumps of period of grating, Figures 10 (d) - (f) show the dependence of period of grating of 1D⊥ structure on voltage $U$ after the exposure time $t_{irr} = 3$, 4 and 8 min ($d/P = 3.2$, 2.5 and 1.3, respectively) for a LC cell consisting of substrates having $W = 1.3 \times 10^{-4}$ J/m$^2$ ($N_{rubb} = 10$) and thickness $d = 20.7$ μm. It should be noted that ratios of $d/P$ for different LC cells, assembled with a pair of substances possessing various values of anchoring energy $W$ are about the same. As was recently theoretically and experimentally shown in [5,7,11,61,62] hysteresis of period of grating can be observed when voltage $U$ is changed in the opposite direction, however, Figure 10 illustrates jumps of period of grating only under the increased voltage. As can be seen in Figure 10, when the ratio $d/P$ is reduced, then the appearance of undulation structure is observed at lower voltage. It is easily seen that for $t_{irr} = 2$ min, $d/P = 3.5$ the appearance of undulation structure is observed at $U_{th} = 4.1$ V (Figure 10 (a)), while for $t_{irr} = 4$ and 6 min, when $d/P = 2.6$ and 1.3, undulations were observed at $U_{th} = 3.4$ and 2.3 V, (Figure 10 (b) and Figure 10 (c), respectively). This regularity is also observed for substrates with various anchoring energy $W$ (at a different $N_{rubb}$), as can be seen from Figure 10.

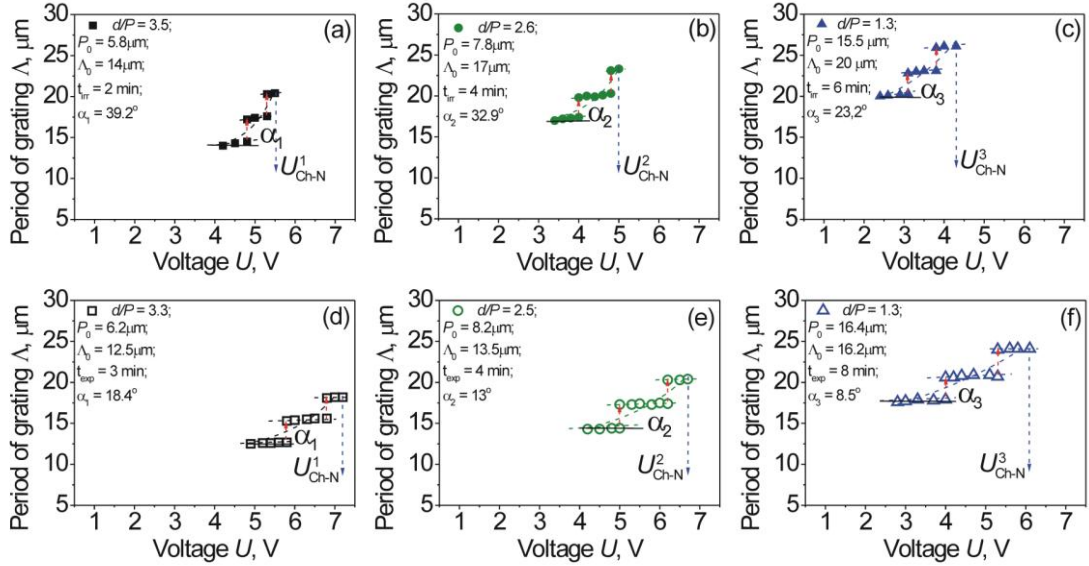

Figure 10. Dependence of period of grating $\Lambda$ of $1D_\parallel$ structure on the applied voltage $U$ after different exposure times $t_{irr}$ for the LC cell, assembled with a pair of substrates possessing anchoring energy $W$: (a), (b) and (c) - $11 \times 10^{-6}$ J/m$^2$ ($N_{rubb} = 1$) and (d), (e) and (f) - $1.3 \times 10^{-4}$ J/m$^2$ ($N_{rubb} = 10$). Jumps of period of grating for 20.3 μm LC cell, when anchoring energy of substrates was $W = 11 \times 10^{-6}$ J/m$^2$, after different exposure times $t_{irr}$: (a) - 2 min, when the ratio $d/P = 3.5$, initial cholesteric pitch $P_0$ and period of grating $\Lambda_0$ are 5.8 and 14 μm, respectively; (b) - 4 min, when the ratio $d/P = 2.6$, initial cholesteric pitch $P_0$ and period of grating $\Lambda_0$ are 7.8 and 17 μm, respectively; (c) - 6 min, when the ratio $d/P = 1.3$, $P_0 = 15.5$ μm and $\Lambda_0 = 20$ μm. The threshold voltage $U^i_{Ch-N}$ of Ch → N transition is: (a) 5.5 V, (b) 5 V and (c) 4.3 V. The tilt angle α of a single domain for different exposure times $t_{irr}$: (a) 39.2°, (b) 32.9° and (c) 23.2°. Jumps of period of grating for 20.7 μm LC cell consisting of a pair of substrates having strong anchoring energy $W = 1.3 \times 10^{-4}$ J/m$^2$ ($N_{rubb} = 10$), after different exposure times $t_{irr}$: (d) - 3 min, when the ratio $d/P = 3.3$, initial cholesteric pitch $P_0$ and period of grating $\Lambda_0$ are 6.2 and 12.5 μm, respectively; (e) - 4 min, when the ratio $d/P = 2.5$, $P_0 = 8.2$ μm and $\Lambda_0 = 13.5$ μm; (f) - 8 min, when the ratio $d/P = 1.3$, $P_0 = 16.4$ μm and $\Lambda_0 = 16.2$ μm. The threshold voltage $U^i_{Ch-N}$ of Ch → N transition is: (d) 7.2 V, (b) 6.7 V and (c) 6.1 V. The tilt angle α of a single domain for different exposure times $t_{irr}$: (a) 18.4°, (b) 13° and (c) 8.5°.

With increase in voltage $U$, jumps of period of grating will be observed and no cholesteric - nematic (Ch → N) transition is occurring. As can be seen in Figure 10 (a), (b) and (c), values of the threshold voltage of Ch → N transition $U^i_{Ch-N} = 5.5$, 5 and 4.3 V (where $i$ = 1, 2 and 3, respectively) depend on exposure times $t_{irr}$ = 2, 4 and 6 min. It is seen that the decrease in ratio $d/P$ (or, in other words, increase in the cholesteric pitch $P$ owing to photoisomerisation of PBM molecules) causes the decrease in $U_{Ch-N}$.

The value of period of grating $\Lambda$ is a bit changing forming a single domain, over a small range of voltage $U$, and further the change in voltage leads to the jump of period of grating, as was recently shown in. [61] However, there is a certain distinction in behavior of a single domain in the case of a different exposure time $t_{irr}$ of a photosensitive cholesteric, namely for a LC cell, assembled with a pair of substrates possessing a certain value of anchoring energy $W$, the line segment of a single domain is inclined with respect to the voltage axis on a certain fixed angle $\alpha_i$ (where $i = 1, 2, 3$), as shown in Figure 10. Here, depending on the duration of UV exposure ($t_{irr}$) the value of a tilt angle of a single domain is changing as $\alpha_1 > \alpha_2 > \alpha_3$, respectively (see Figure 10 (a), Figure 10 (b) and Figure 10 (c)). As was recently shown in [61], the tilt angel of a single domain depends on the value of anchoring energy $W$ of substrates.

In this work it was also found that the value of the tilt angle $\alpha$ also depends on the exposure time $t_{irr}$ at fixed anchoring energy $W$, as shown in Figure 11. If anchoring energy $W$ is strong (for example, at $N_{rubb} = 10$, $W = 1.3 \times 10^{-4}$ J/m$^2$), then the tilt angle $\alpha_i$ (where $i = 1, 2, 3,...$) is less than $\alpha_i$ in the case of weak anchoring energy $W$ (for example, at $N_{rubb} = 1, 5$ and $25$ from the Table 1 the anchoring energy is $W = 11 \times 10^{-6}$, $17 \times 10^{-6}$ and $18 \times 10^{-6}$ J/m$^2$, respectively), as can be seen in Figure 11. The increase in the cholesteric pitch $P$ (or, in other words, the decrease in the helical twisting power $\beta$ due to *trans-cis* photoisomerisation of chiral molecules PBM) leads to the decrease in the contribution of chirality under a competition between anchoring energy $W$ and chiral power $\beta$. As a result, molecules of the liquid crystal interact more strongly with the surface of alignment layers of the LC cell, and the tilt angle $\alpha$ is decreasing.

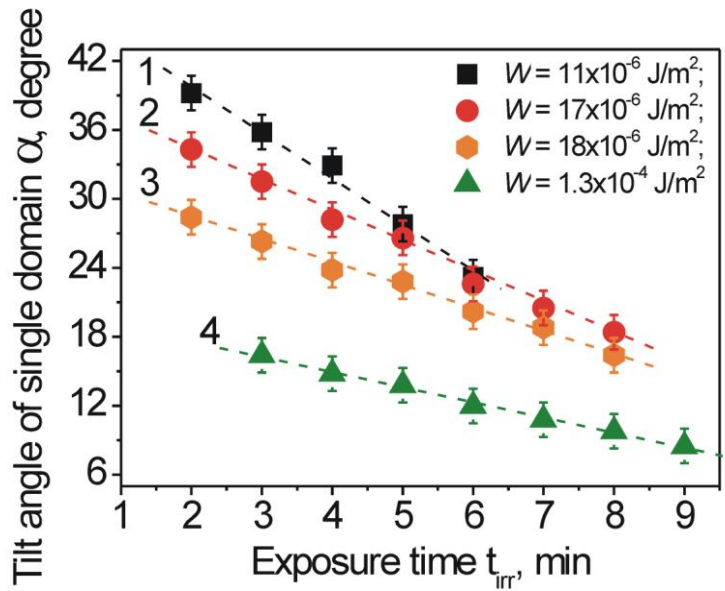

Figure 11. Dependence of the tilt angle of a single domain $\alpha$ on the exposure time $t_{irr}$, when the threshold voltage $U_{th}$ were applied to the LC cells, assembled with a pair of substrates possessing various values of anchoring energy $W$: $11 \times 10^{-6}$ J/m$^2$ (square

symbols, line 1), $17 \times 10^{-6}$ J/m$^2$ (solid circles, line 2), $18 \times 10^{-6}$ J/m$^2$ (solid diamonds, line 3) and $1.3 \times 10^{-4}$ J/m$^2$ (solid triangles, line 4).

### *3.5. Diffraction efficiency of 1D$_\parallel$ cholesteric grating*

It is should be noted that the diffraction efficiency (DE) of different types of cholesteric gratings (2D and 1D) was in detail measured in. [29-31,37,46,63] This section will describe only 1D$_\parallel$ cholesteric diffraction grating with an electrically controlled period $\Lambda$, which is observed under conditions $1.2 \leq d/P \leq 3.5$.

An experimental setup for the measurement of the intensity of the diffracted beam $I_m$ is shown in Figure 12. Output monochromatic linear-polarised light from He-Ne laser illuminates the LC cell, where under alternating sinusoidal voltage, 1D$_\parallel$ undulation structure was formed. An iris diaphragm was placed behind the LC cell in order to select certain diffraction orders $m$. The total intensity $I_0$ of the beam on PD was measured without applied voltage $U = 0$ V. The intensity of the diffracted order $I_m$ was measured under the condition of the applied voltage to the LC cell, and there is $U \geq U_{th}$ (where $U_{th}$ is threshold voltage, when 1D$_\parallel$ undulation structure is appeared).

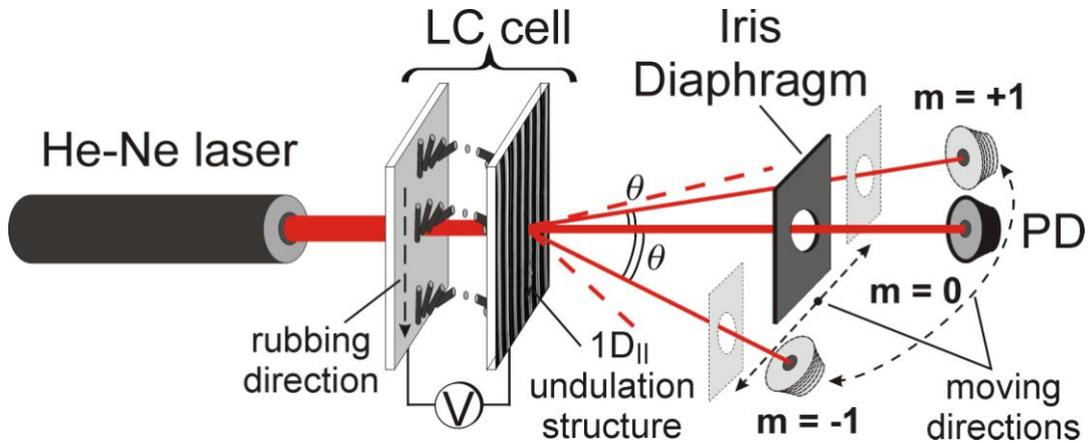

Figure 12. Experimental setup for the observation of diffraction and measurement of the diffraction efficiency. The iris diaphragm can be moved to select the diffraction order $(0, \pm 1, \pm 2, \textit{etc.})$

For the normal incident beam, the directions (or angles) of the diffraction orders $m = 0; \pm 1; \pm 2; \textit{etc.}$, are: [40-42]

$$\theta_m = \arcsin(m \cdot \lambda / \Lambda) \qquad (3).$$

In Figure 13 the direction of the second diffraction maximum (angle $\theta_m$, $m = 2$) as the function of the exposure time $t_{irr}$ for the LC cell consisting of a pair of substrates possessing a certain value of anchoring energy $W$ is shown. The increase in the exposure time $t_{irr}$ causes the increase in period of grating $\Lambda$, owing to *trans-cis* photoisomerisation of PBM molecules, and the decrease in the diffraction angle $\theta_m$ (for instance $m = 2$) is observed. As can be seen from Figure 13, the speed of the decrease in the diffraction angle $\theta_m$ is more for LC cells, assembled with a pair of substrates having weak anchoring energy, than in the case of strong anchoring energy $W$.

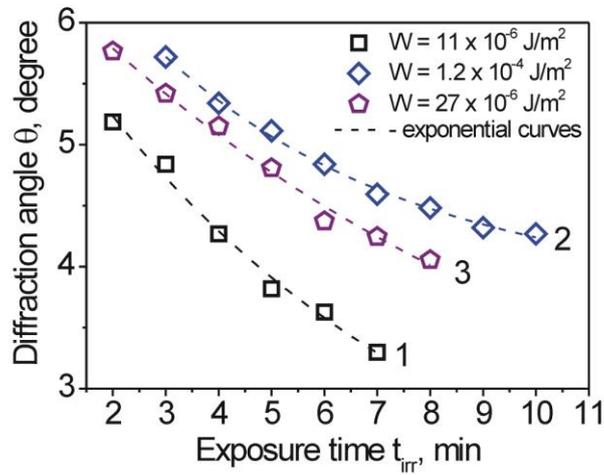

Figure 13. Dependence of the direction of the 2$^{nd}$ diffraction maximum (diffraction angle $\theta_m$, $m = 2$) on the exposure time $t_{irr}$ of the LC cell, assembled with a pair of substrates having anchoring energy $W$: (curve 1, open squares) – $11 \times 10^{-6}$ J/m$^2$ ($N_{rubb} = 1$); (curve 2, open diamonds) – $1.2 \times 10^{-4}$ J/m$^2$ ($N_{rubb} = 15$) and (curve 3, open pentagons) – $27 \times 10^{-6}$ J/m$^2$ ($N_{rubb} = 20$). The thickness of the LC cells was $d = 20.2$ μm (curve 1) and $d = 20.3$ μm (curve 2 and curve 3).

As can be seen from Figure 10, the increase in the applied voltage $U$ gives rise to the increase in period of grating $\Lambda$ of 1D$_{\parallel}$ undulation structure. In this case, as can be seen from Figure 14, the decrease in the angles $\theta_m$ (for instance $m = \pm 2$) of diffraction orders is observed. As in the case of the jump-wise changing of period of grating under an applied voltage $U$ to the LC cell (Figure 10), the calculated, by using Equation (3), angle of diffraction $\theta_m$ ($m = \pm 2$) also changes by jumps as shown in Figure 14, by using symbols (open and solid squares, circles and triangles). The dependencies of the calculated and measured (by using diffraction pattern (solid and open stars)) angles of diffraction $\theta_m$ ($m = \pm 2$) on the applied voltage $U$ for different LC cells assembled with a pair of ODAPI substrates having weak anchoring energy $W = 11 \times 10^{-6}$ J/m$^2$ ($N_{rubb} = 1$) and strong $W = 1.3 \times 10^{-4}$ J/m$^2$ ($N_{rubb} = 10$) at a certain fixed exposure time $t_{irr}$ (curves 1, 2 and 3) are shown in Figure 14 (a) and Figure 14 (b), respectively.

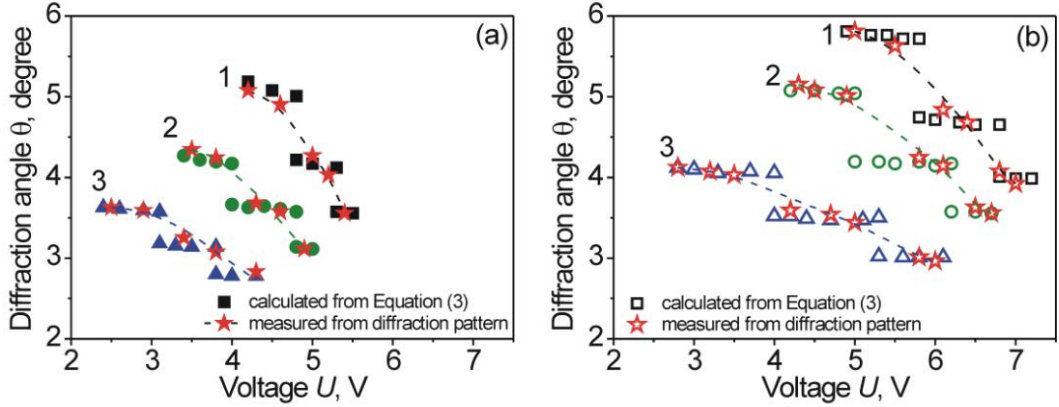

Figure 14. Dependence of the diffraction angle $\theta_m$ ($m = \pm 2$) on the applied voltage $U$ after different exposure times $t_{irr}$ for the LC cell, assembled with substrates having: (a) weak anchoring energy $W = 11 \times 10^{-6}$ J/m$^2$ ($N_{rubb} = 1$) and (b) strong $W = 1.3 \times 10^{-4}$ J/m$^2$ ($N_{rubb} = 10$). (a) Jumps of the calculated diffraction angle $\theta_m$ for 20.3 μm LC cell, when anchoring energy of substrates was $W = 11 \times 10^{-6}$ J/m$^2$, after different exposure times $t_{irr}$: (curve 1, solid squares) - 2 min, when the ratio $d/P = 3.5$, initial cholesteric pitch $P_0$ and period of grating $\Lambda_0$ are 5.8 and 14 μm, respectively; (curve 2, solid circles) - 4 min, when the ratio $d/P = 2.6$, initial cholesteric pitch $P_0$ and period of grating $\Lambda_0$ are 7.8 and 17 μm, respectively; (curve 3, solid triangles) - 6 min, when the ratio $d/P = 1.3$, $P_0 = 15.5$ μm and $\Lambda_0 = 20$ μm. (b) Jumps of the calculated diffraction angle $\theta_m$ for 20.7 μm LC cell consisting of substrates having strong anchoring energy $W = 1.3 \times 10^{-4}$ J/m$^2$, after different exposure times $t_{irr}$: (curve 1, open squares) - 3 min, when the ratio $d/P = 3.3$, initial cholesteric pitch $P_0$ and period of grating $\Lambda_0$ are 6.2 and 12.5 μm, respectively; (curve 2, open circles) - 4 min, when the ratio $d/P = 2.5$, $P_0 = 8.2$ μm and $\Lambda_0 = 13.5$ μm; (curve 3, open triangles) - 8 min, when the ratio $d/P = 1.3$, $P_0 = 16.4$ μm and $\Lambda_0 = 16.2$ μm. The measured diffraction angle $\theta_m$ from the diffraction pattern of the LC cell with: (a) weak anchoring energy $W = 11 \times 10^{-6}$ J/m$^2$ (solid stars) and (b) strong anchoring energy $W = 1.3 \times 10^{-4}$ J/m$^2$ (open stars).

The diffraction efficiency (DE) was calculated with the intensity $I_m$ of diffracted order $m$ and the total intensity $I_0$, by using equation: [40,41]

$$\eta_m(U) = \frac{I_m(U)}{I_0} \tag{4}$$

The DE of 1D$_\parallel$ grating (for diffraction orders $m = 0, 1, 2$) as the function of voltage $U$ applied to the LC cell with a pair of substrates having weak (open symbols) and strong (solid symbols) anchoring energy $W$ is shown in Figure 15, when cholesteric mixtures were UV illuminated during the same exposure, for instance $t_{irr} = 4$ min.

The rise of the voltage $U$ results in the decrease in DE of 0 - order and the diffracted intensity for 1 - order is increasing. As can be seen from Figure 15, the values of DE depend on voltage $U$. The range of the controlling voltage $U$ is large for the LC cell consisting of a pair of substrates having strong anchoring energy $W$ and *vice versa*.

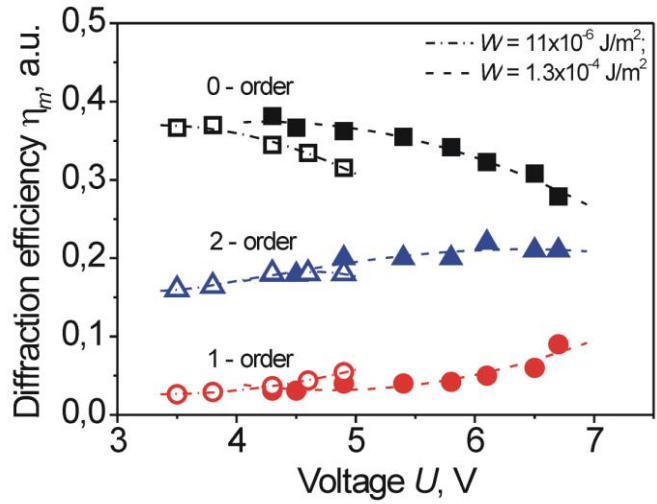

Figure 15. Diffraction efficiency of 1D$_\parallel$ grating (0 -, 1 -, and 2 - orders) as the function of the applied voltage $U$. The LC cells were assembled with a pair of substrates possessing various values of anchoring energy: (open symbols) – weak $W = 11 \times 10^{-6}$ J/m$^2$ ($N_{rubb} = 1$) and (solid symbols) – strong $W = 1.3 \times 10^{-4}$ J/m$^2$ ($N_{rubb} = 10$). The thickness of the LC cell was $d = 20.3$ μm (open symbols) and 20.7 μm (solid symbols). The cholesteric mixture was illuminated during $t_{irr} = 4$ min. Symbols represent experimental data, and curves are the simulation results.

By comparing values of DE $\eta_m$ from Figure 15 measured in, [24,29,33,63,64] it can be concluded that results are very similar, though obtained values of DE are a little lower. It is obviously that one possible reason for lower values $\eta_m$ is that the appearance of light scattering around diffraction orders (the so-called arc-shaped diffraction orders) [29] is observed on the defects of 1D$_\parallel$ undulation structure during the increase in voltage $U$.

**Conclusions**

In this work the processes of electro- and photoswitching of undulation structure, that can be useful for a creation of cholesteric gratings, were studied. The influence of anchoring energy $W$ (or $N_{rubb}$), the exposure time $t_{irr}$ and voltage of the applied field $U$ on main features of diffraction gratings were investigated. It was shown that due to the usage of photosensitive CLC, the changing ratio $d/P$ during UV exposure (for the LC cell with a fixed thickness $d$) leads to sequential transitions of 2D, 1D$_\parallel$ and 1D$_\perp$ undulation structure. In addition, it was shown that due to *trans-cis* photoisomerisation of PBM molecules during UV exposure, the various initial period of grating $\Lambda$, appearing in the alternating electric field, could be obtained. Depending on the value of anchoring energy $W$ ($N_{rubb}$), period of grating $\Lambda$ can be changed by jumps, when 1D$_\parallel$ cholesteric grating is formed under the applied to the LC cell voltage. Detail studies of characteristics of the CLC grating, namely the diffraction angle $\theta_m$ and DE $\eta_m$ of 1D$_\parallel$ undulation structure as an electro- and photocontrolled diffraction grating, were carried out. However, as it was experimentally obtained, DE $\eta_m$ is less than for the CLC grating studied in. [29-31,38,46] Although the photosensitive ChD, used in this manuscript, has two isomers with similar spectral characteristics, [49,50] here the main goal was to demonstrate the possible usage of the photosensitive CLC with the variable helix pitch $P$ for electro- and photoswitchable diffraction gratings and their switching between 2D and 1D (1D$_\parallel$ and 1D$_\perp$) types for the LC cell with a fixed thickness $d$. In the case of the usage of photoreversible ChDs with absorption spectra of their isomers in different wavelength ranges, for instance as was described in, [13-17] the reversible photoswitching between undulation structures (or, in other words, the type of cholesteric gratings) could be used for different optical applications.

Acknowledgements

Author I.G. thanks W. Becker (Merck) for his generous gift of the nematic E7, Prof. I. Gerus (Institute of Bio-organic Chemistry and Petrochemistry, NAS of Ukraine) for the kind provision of polymers, Scientific researcher P. Tytarenko (V. Ye. Lashkaryov Institute of Semiconductor Physics, NAS of Ukraine) for the preparation of indium-tin oxide (ITO)-coated on glass substrates, Field service specialist IV V. Danylyuk (Dish LLC, USA) for his gift of some Laboratory equipments and Dr. S. Lukyanets (Institute of Physics, NAS of Ukraine) for the helpful discussions.

References:

[1]. de Gennes PG, Prost J. The physics of liquid crystals. Oxford: Clarendon Press; 1993.

[2]. Chilaya GS, Lisetskii LN. Helical twist in cholesteric mesophases. Sov Phys Usp. 1981; 24(*6*): 496-510.

[3]. Zeldovich BY, Tabiryan NV. Equilibrium structure of a cholesteric with homeotropic orientation on the walls. JETP. 1982; 56(*3*): 563-566.

[4]. Kimura H, Hosino M, Nakano H. Temperature dependent pitch in cholesteric phase. J Phys Colloques. 1979; 40:C3-174-C3-177. doi: 10.1051/jphyscol:1979335

[5]. Belyakov VA, Kats EI. Surface anchoring and temperature variations of the pitch in thin cholesteric layers. JETP. 2000; 91(*3*): 488-491.

[6]. Chilaya G, Hauck G, Koswig HD, et al. Field induced increase of pitch in planar cholesteric liquid crystals. Cryst Res Technol. 1997; 32(*3*): 401–405. doi: 10.1002/crat.2170320306

[7]. Belyakov A. Untwisting of the helical structure in a plane of chiral liquid crystals. JETP Letters. 2002; 72(2): 88-92.

[8]. Greubel W, Wolff U, Krüger H. Electric field induced texture changes in certain nematic/cholesteric liquid crystal mixtures. Mol Cryst Liq Cryst. 1973; 24 (1-2): 103-111. http://dx.doi.org/10.1080/15421407308083392

[9]. Xiang J, Li Y, Paterson DA, et al. Electrically tunable selective reflection of light from ultraviolet to visible and infrared by heliconical cholesterics. Adv Mater. 2015; 27(19): 3014-3018. doi: 10.1002/adma.201500340

[10]. Salili SM, Ribeiro de Almeida RR, Challa PK et al. Spontaneously modulated chiral nematic structures of flexible bent-core liquid crystal dimers. Liq Cryst. 2017; 44(*1*): 160-167. http://dx.doi.org/10.1080/02678292.2016.1225836

[11]. Kiselev AD, Sluckin TJ. Twist of cholesteric liquid crystal cells: stability of helical structures and anchoring energy effects. Phys Rev E. 2005; 71:031704–15. doi:10.1103/PhysRevE.71.031704

[12]. Ishikawa T, Lavrentovich OD. Defects in liquid crystals: computer simulations, theory and experiments. Netherlands: Kluwer Academic Publishers; 2001.

[13]. Denekamp C, Feringa BL. Optically active diarylethenes for multimode photoswitching between liquid crystalline phases. Adv Mater. 1998; 10(*14*): 1080–1082. doi:10.1002/(SICI)1521-4095(199810)10:14<1080::AID-ADMA1080>3.0.CO;2-T

[14]. Bosco A, Jongejan MGM; Eelkema R, et al. Photoinduced reorganization of motor-doped chiral liquid crystals: bridging molecular isomerization and texture rotation. J Am Chem Soc. 2008; 130: 14615-14624. doi: 10.1021/ja8039629

[15]. Eelkema R, Feringa BL. Amplification of chirality in liquid crystals. Org Biomol Chem. 2006; 4, 3729-3745. doi: 10.1039/B608749C

[16]. Wang Y, Urbas A, Li Q. Reversible visible-light tuning of self-organized helical superstructures enabled by unprecedented light-driven axially chiral molecular switches. J Am Chem Soc. 2012; 134: 3342-3345. doi: 10.1021/ja211837f

[17]. Wang Y, Li Q. Light-driven chiral molecular switches or motors in liquid crystals. Adv Mater. 2012; 24 (*15*): 1926-1945. doi: 10.1002/adma.201200241

[18]. Terenetskaya I, Gvozdovsky I. Development of personal UV biodosimeter based on vitamin D photosynthesis. Mol Cryst Liq Cryst. 2001; 368: 551-558. http://dx.doi.org/10.1080/10587250108029987

[19]. Aronishidze M, Chanishvili A, Chilaya G, et al. Color change effect based on provitamin D phototransformation in cholesteric liquid crystalline mixtures. Mol Cryst Liq Cryst. 2004; 420: 47-53. http://dx.doi.org/10.1080/15421400490478353

[20]. Gvozdovskyy I, Yaroshchuk O, Serbina M. Light-induced nematic - cholesteric structural transitions in the LC cells with homeotropic anchoring. Mol Cryst Liq Cryst. 2011; 546: 202/[1672] - 208/[1678]. http://dx.doi.org/10.1080/15421406.2011.571161

[21]. Ilchishin IP, Lisetskiy LM, Mykytiuk TV, et al. Reversible phototuningof lasing frequency in a due-doped cholesteric liquid crystal. Ukr J Phys. 2011; 56(*4*): 333-338.

[22]. Schadt M, Helfrich W. Voltage-dependent optical activity of a twisted nematic liquid crystal. Appl Phys Lett. 1971; 18 (*4*): 127-128. doi: 10.1063/1.1653593

[23]. Crandall KA, Fisch MR, Petschek RG, et al. Vanishing Freedericksz transition threshold voltage in a chiral nematic liquid crystal. Appl Phys Lett. 1994; 64: 1741-1743. doi: 10.1063/1.111796

[24]. Varanytsia A, Chien L-C. A spatial light modulator with two-dimentional array of liquid crystal bubbles. SID 14 Digest. 2014; 45:1492-1495.

[25]. Meyer RB, Lonberg F, Chang CC. Cholesteric liquid crystal smart reflectors. Mol Cryst Liq Cryst. 1996; 288 (*1*): 47-61. http://dx.doi.org/10.1080/10587259608034583

[26]. Drolet JJP, Chuang E, Barbastathis G, et al. Compact, integrated dynamic holographic memory with refreshed holograms. Opt Lett. 1997; 22 (*8*): 552-554. doi: 10.1364/OL.22.000552

[27]. Lu MH. Bistable reflective cholesteric liquid crystal display. J Appl Phys. 1997; 81(3): 1063-1066. http://dx.doi.org/10.1063/1.363867

[28]. Faklis D, Morris GM. Diffractive optics technology for display applications. In: Wu MH, editor. Light Source, Optics, and Other Critical Components II. Projection Displays. Proceedings SPIE 2407; 1995 February 5; San Jose (CA); 1995. doi: 10.1117/12.205913

[29]. Subacius D, Shiyanovskii SV, Bos Ph, et al. Cholesteric gratings with field-controlled period. Appl Phys Lett. 1997; 71(*23*): 3323-3325.

[30]. Senyuk B, Smalyukh I, Lavrentovich O. Elecrtically-controlled two-dimensional gratings based on layers undulationsin cholesteric liquid crystals. In: Khoo I-C, editor. Liquid Crystals IX. IR, microwave, and beam steering. Proceedings of SPIE 59360W; 2005 August 20; San Diego. California (USA); 2005. doi:10.1117/12.615976

[31]. Senyuk BI, Smalyukh II, Lavrentovich OD. Switchable two-dimensional gratings based on field-induced layer undulations in cholesteric liquid crystals. Opt Lett. 2005; 30(*4*): 349-351. doi: 10.1364/OL.30.000349

[32]. Lin C-H, Chiang R-H, Liu S-H, et al. Rotatable diffractive grating based on hybrid-aligned cholesteric liquid crystals. Opt Express. 2012; 20(24): 26837-26844. doi:10.1364/OE.20.026837

[33]. Ryabchun A, Bobrovsky A, Stumpe J, et al. Electroinduced diffraction gratings in cholesteric polymer with phototunable helix pitch. Adv Optical Mater. 2015; 3(*10*): 1462-1469. doi: 10.1002/adom.201500293

[34]. Ryabchun A, Bobrovsky A, Stumpe J, et al. Rotatable diffraction gratings based on cholesteric liquid crystals with phototunable helix pitch. Adv Optical Mater. 2015; 3(9): 1273-1279. doi: 10.1002/adom.201500159

[35]. Gvozdovskyy I, Yaroshchuk O, Serbina M, et al. Photoinduced helical inversion in cholesteric liquid crystal cells with homeotropic anchoring. Opt Express. 2012; 20(*4*): 3499-3508. doi: 10.1364/OE.20.003499

[36]. Li W-S, Shen Y, Chen Z-J, et al. Demonstration of patterned polymer-stabilized cholesteric liquid crystal textures for anti-counterfeiting two-dimensional barcodes. Appl Opt. 2017; 56(*3*): 601-606. doi: 10.1364/AO.56.000601

[37]. Hamdi R, Petriashvili G, De Santo MP, et al. Electrically controlled 1D and 2D cholesteric liquid crystal gratings. Mol Cryst Liq Cryst. 2012; 553: 97-102. http://dx.doi.org/10.1080/15421406.2011.609436

[38]. Fuh AY-G, Lin Ch-H, Huang Ch-Y. Dynamic pattern formation and beam-steering characteristics of cholesteric gratings. Jpn J Appl Phys 2002; 41(*1*): 211-218. doi: 10.1143/JJAP.41.211

[39]. Czajkovski M, Klajn J, Cybińska J, et al. Cholesteric gratings induced by electric field in mixtures of liquid crystal and novel chiral ionic liquid. Liq Cryst. 2016; http://dx.doi.org/10.1080/02678292.2016.1254825

[40]. Klein WR, Cook BD. Unified approach to ultrasonic light diffraction. IEEE Trans Sonics Ultrason. 1967; SU-14(*3*):123–134.

[41]. Moharam MG, Young L. Criterion for Bragg and Raman-Nath diffraction regimes. Appl Opt. 1978; 17(*11*): 1757-1759. doi: 10.1364/AO.17.001757

[42]. Gaylord TK, Moharam MG. Thin and thick gratings: terminology clarification. Appl Opt. 1981; 20: 3271-3273. doi: 10.1364/AO.20.003271

[43]. Helfrich W. Deformation of cholesteric liquid crystals with low threshold voltage. Appl Phys Lett. 1970; 17: 531-532. doi: 10.1063/1.1653297

[44]. Helfrich W. Electrohydrodynamic and dielectric instabilities of cholesteric liquid crystals. J Chem Phys. 1971; 55: 839-842. http://dx.doi.org/10.1063/1.1676151

[45]. Varanytsia A, Chien L-C. Photoswitchable and dye-doped bubble domain texture of cholesteric liquid crystals. Opt Lett. 2015; 40(*19*): 4392-4395. doi: 10.1364/OL.40.004392

[46]. Subacius D, Bos Ph, Lavrentovich OD. Switchable diffractive cholesteric gratings. Appl Phys Lett. 1997; 71(*10*): 1350-1352. doi: 10.1063/1.119890

[47]. Gerritsma CJ, van Zanten P. Periodic perturbulations in the cholesteric plane texture. Phys Lett. 1971; 37A(*1*): 47-48. https://doi.org/10.1016/0375-9601(71)90325-2

[48]. Senyuk BI, Smalyukh II, Lavrentovich OD. Undulations of lamellar liquid crystals in cells with finite surface anchoring near and well above the threshold. Phys Rev E. 2006; 74: 011712-1-13. doi: 10.1103/PhysRevE.74.011712

[49]. Kutulya LA, Kuz'min VE, Stel'makh IB, et al. Quantitative aspects of chirality. III. Description of the influence of the structure of chiral compounds on their twisting power in the nematic mesophase by means of the dissymmetry function. J Phys Org Chem. 1992; 5: 308-316. doi: 10.1002/poc.610050605

[50]. Yarmolenko SN, Kutulya LA, Vaschenko VV, et al. Photosensitive chiral dopants with high twisting power. Liq Cryst. 1994; 16(*5*): 877-882. doi:10.1080/02678299408027858

[51]. Licristal brochure, Merck Liquid crystals, 1994.

[52]. Rynes EP, Brown CV, Strömer JF. Method for the measurement of the $K_{22}$ nematic elastic constant. App Phys Lett. 2003; 82(*1*): 13–15. doi:10.1063/1.1534942

[53]. Yang F, Sambles JR, Bradberry GW. Half-leaky guided wave determination of azhimuthal anchoring energy and twist elastic constant of a homogeneously aligned nematic liquid crystal. J Appl Phys. 1999; 85: 728–733. doi:10.1063/1.369153

[54]. Grandjean F. Existence des plans differences équidistants normal a l'axe optique dans les liquides anisotropes [Existence of the optical axis planes equidistant normal differences in anisotropic fluids]. C R Hebd Seances Acad. 1921; 172: 71-74. French.

[55]. Cano R. Interprétation des discontinuités de Grandjean [Interpretation of discontinuities Grandjean]. Bull Soc Fr Mineral Crystalogr. 1968; 91: 20-27. French.

[56]. Andrienko D, Kurioz Y, Nishikawa M, et al. Control of the anchoring energy of rubbed polyimide layers by irradiation with depolarized UV light. Jpn J Appl Phys. 2000; 39(Part 1, No. 3A): 1217–1220. doi:10.1143/JJAP.39.1217

[57]. Smalyukh II, Lavrentovich OD. Anchoring-mediated interaction of edge dislocations with bounding surfaces in confined cholesteric liquid crystals. Phys Rev Lett. 2003; 90(*8*): 085503-1-4. doi: 10.1103/PhysRevLett.90.085503

[58]. Lavrentovich OD, Yang D-K. Cholesteric cellular patterns with electric-field-controlled line tension. Phys Rev E. 1998; 57:R6269-R6272. doi: 10.1103/PhysRevE.57.R6269

[59]. Gerus I, Glushchenko A, Kwon S-B, et al. Anchoring of a liquid crystal on photoaligning layer with varying surface morphology. Liq Cryst. 2001; 28: 1709-1713. http://dx.doi.org/10.1080/02678290110076371

[60]. Andrienko D, Dyadyusha A, Iljin A, et al. Measurement of azimuthal anchoring energy of nematic liquid crystal on photoaligning polymer surface. Mol Cryst Liq Cryst. 1998; 321:271-281. http://dx.doi.org/10.1080/10587259808025093

[61]. Gvozdovskyy I. Influence of the anchoring energy on jumps of the period of stripes in thin planar cholesteric layers under the alternating electric field. Liquid Crystals. 2015; 41(*10*): 1495-1504. doi: 10.1080/02678292.2014.927930

[62]. McKay G. Bistable surface anchoring and hysteresis of pitch jumps in a planar cholesteric liquid crystal. Eur Phys J E. 2012; 35(*8*): 74-1-8. doi: 10.1140/epje/i2012-12074-1

[63]. Jepsen ML, Gerritsen HJ. Liquid-crystal-filled gratings with diffraction efficiency. Opt Lett. 1996; 21(*14*): 1081-1083. doi: 10.1364/OL.21.001081

[64]. Varanytsia A, Chein L-C. Bistable and photo switchable liquid crystal diffractive grating. Poster session presented at: 25th International Liquid Crystal Conference; 2014 Jun 29 – Jul 4; Dublin, Ireland.